# Search for *Evergreens* in Science: A Functional Data Analysis[*]

## Ruizhi Zhang[1], Jian Wang[2,3] & Yajun Mei[1]

[1]H. Milton Stewart School of Industrial & Systems Engineering, Georgia Institute of Technology

[2]Center for R&D Monitoring and Department of Managerial Economics, Strategy & Innovation, KU Leuven

[3]German Center for Higher Education Research and Science Studies, DZHW Berlin

Emails: zrz123@gatech.edu, jian.wang@kuleuven.be, ymei@isye.gatech.edu

## May 17, 2017

### Abstract

*Evergreens* in science are papers that display a continual rise in annual citations without decline, at least within a sufficiently long time period. Aiming to better understand evergreens in particular and patterns of citation trajectory in general, this paper develops a functional data analysis method to cluster citation trajectories of a sample of 1699 research papers published in 1980 in the American Physical Society (APS) journals. We propose a functional Poisson regression model for individual papers citation trajectories, and fit the model to the observed 30-year citations of individual papers by functional principal component analysis and maximum likelihood estimation. Based on the estimated paper-specific coefficients, we apply the K-means clustering algorithm to cluster papers into different groups, for uncovering general types of citation trajectories. The result demonstrates the existence of an evergreen cluster of papers that do not exhibit any decline in annual citations over 30 years.

[*]Ruizhi Zhang, Jian Wang & Yajun Mei. (2017). Search for evergreens in science: A functional data analysis. *Journal of Informetrics*, 11(3), 629–644. http://dx.doi.org/10.1016/j.joi.2017.05.007 ©2017 Elsevier Ltd.

The authors thank the editor and three anonymous referees for their constructive comments which have substantially improved this paper. R. Zhang and Y. Mei were supported in part by the NSF grant CMMI-1362876, and J. Wang by a postdoctoral fellowship from the Research Foundation – Flanders (FWO). Data used in this paper are from a bibliometric database developed by the Competence Center for Bibliometrics for the German Science System (KB) and derived from the 1980 to 2012 Science Citation Index Expanded (SCI-E), Social Sciences Citation Index (SSCI), Arts and Humanities Citation Index (AHCI), Conference Proceedings Citation Index–Science (CPCI-S), and Conference Proceedings Citation Index–Social Science & Humanities (CPCI-SSH) prepared by Thomson Reuters (Scientific) Inc. (TR®), Philadelphia, Pennsylvania, USA: ©Copyright Thomson Reuters (Scientific) 2013. KB is funded by the German Federal Ministry of Education and Research (BMBF, project number: 01PQ08004A).

# 1. Introduction

Science is a skewed world where a small number of publications receive a disproportionate amount of citations. What do citation trajectories of the most cited papers look like? Do they follow the "typical" citation trajectory documented in the literature, specifically, the annual citations of a paper rise to a peak in the first few years after publication and then slowly fade away over time? *Fig. 1* plots annual citations of the top ten most cited papers published in the American Physical Society (APS) journals, and their annual citations are counted in the Web of Science (WoS) from the year of publication to 2016. Among them the youngest was published in 1999, and the oldest 1964. Correspondingly, the length of their observed citation trajectories range from 18 to 53 years (*Appendix A*). In addition to their exceptionally large number of citations, a remarkable observation is that most of them (at least seven out of ten) do not even show any sign that their annual citations are about to peak and will start to decline in the near future. We refer to this phenomenon of continual rise in annual citations without decline as *evergreens*, which clearly violates the "typical" pattern of citation trajectory. Although we cannot predict whether these papers will remain highly cited in the future, the fact that they have not yet become obsolete after up to 53 years calls for attention, especially considering that the majority of papers reach their citation peak around the $3^{rd}$ or $5^{th}$ year after publication and that most bibliometric analyses examine citations in a relatively short time window.

Insert Fig. 1 here

The objective of this paper is to better understand *evergreens* in particular and patterns of citation trajectory in general. Moreover, do *evergreens* constitute a general type of citation trajectory, or are they so rare that they cannot be captured in any statistical cluster analysis?

To this end, we develop a functional data analysis (FDA) method to analyze the 30-year citation trajectories of a sample of publications published in 1980 in APS journals. Our FDA method integrates functional principal components analysis, Poisson regression, and $K$-means clustering. More specifically, we model the citation trajectories of individual publications by a small number of common basis functions and paper-specific coefficients on these basis functions. For each paper, its 30-dimensional vector of citations can be characterized by its coefficients on the common basis functions, which subsequently serve as inputs for the $K$-means clustering, to uncover general types of citation trajectories. Results of our cluster analysis provide strong evidence that *evergreens* exist as a general class of citation trajectory. In addition, we are not able to predict whether a paper will become an *evergreen* by some *ex ante* paper features such as the number of authors and references.

The remainder of this paper is organized as follows. We begin with a brief review of previous cluster analyses of citation trajectories, as well as the method of functional data analysis, followed by a description of our dataset. Next, our proposed model and method is presented, with the emphasis on how to combine functional principal component analysis, Poisson regression, and $K$-means clustering algorithm for modeling and clustering citation trajectories. Then we report the empirical results of our proposed model and method to the real citation dataset. Implications of our findings are also discussed.

## 2. Prior literature

### 2.1. Clustering citation trajectories

Citation ageing has been a long-standing research topic, and different patterns of citation trajectories have been documented in the bibliometrics literature (Avramescu, 1979; Garfield,

1980; Aversa, 1985; Line, 1993; Glänzel & Schoepflin, 1995; Redner, 2005; Rogers, 2010; Wang, 2013; Baumgartner & Leydesdorff, 2014). Aversa (1985) conducted probably the first rigorous statistical analysis of citation trajectories, investigating 9-year citation trajectories of 400 highly cited papers published in 1972 and applying the *K*-means clustering algorithm to the normalized annual citation counts (i.e., annual citations divided by total citations in the whole studied time period). Aversa (1985) identified two clusters: *delayed rise - slow decline* and *early rise - rapid decline*.

Costas, van Leeuwen, and van Raan (2010) analyzed about 30 million documents in WoS published between 1980 and 2008. Following Price's observation, documented in his personal communication to Aversa (1985), Costas et al. (2010) classified papers into three categories: 50% papers as *normal documents*, 25% as *delayed documents*, and 25% as *flashes-in-the-pan*. However, these three clusters are defined based on a single real-valued summary statistics of individual papers, *Year 50%*, defined as the year when a paper has cumulated half of its total citations up to year 2008. In addition, there is no statistical justification on the proportion of these three clusters.

More recently, Colavizza and Franceschet (2016) examined about half million papers published in APS journals and applied the spectral clustering method on the normalized annual citations received by these papers within the APS database. The three identified general types of citation trajectories are *middle-of-the-roads*, *sprinters,* and *marathoners*. *Middle-of-the-roads* papers display an average citation ageing pattern, and can be viewed as corresponding to *normal documents*. *Sprinters* have an early and high peak and a fast decline, which can be viewed as *flashes-in-the-pan*. *Marathoners* represent "fast or slow-rise, moderately peaked histories, followed by a slow decline, or absence of decline, or even a constant rise in received citations over time" and therefore can correspond to *delayed documents* or *evergreens*.

The phenomenon of *evergreens*, which were emphasized by Avramescu (1979) and Price (see Aversa (1985)), were not identified by clustering analyses in Aversa (1985) and Costas et al. (2010), while *marathoners* in some specifications in Colavizza and Franceschet (2016) also display a continually increasing annual citation curve.

## 2.2. *Functional data analysis*

Functional data analysis (FDA) is a recent new development in the field of statistics and has a tremendous growth over the past decades (Besse & Ramsay, 1986; Rice & Silverman, 1991; Hoover, Rice, Wu, & Yang, 1998; Ramsay & Silverman, 2005; Yao, Müller, & Wang, 2005; Hall, Müller, & Wang, 2006; Leng & Müller, 2006; Hadjipantelis, Aston, & Evans, 2012). FDA might be particularly useful for bibliometric analysis for two reasons: First, FDA is a non-parametric method and is therefore useful for analyzing bibliometric data for which the underlying distribution is often unclear. Second, FDA analyzes high-dimensional data, such as curves and shapes, which are of particular interest to bibliometric studies. Using regression analysis as an analogy, while traditional regression analysis only allows one real-valued dependent variable, FDA allows both dependent and independent variables to be multidimensional.

Most FDA methods deal with continuous data, but paper citations to be analyzed in this study are discrete count data. There are only a few FDA studies dealing with count variables (van der Linde, 2009; Serban, Staicu, & Carroll, 2013; Wu, Müller, & Zhang, 2013), and our proposed method is different from these limited existing FDA methods for Poisson data. Specifically, we adapt the methods in Rice and Silverman (1991) from Gaussian distributed data to Poisson count data by exploring the close relationship between Poisson and Gaussian distributions.

## 3. Data

The data used for this study are research papers published in 1980 in the American Physical Society (APS) journals, specifically six journals which were active in 1980: *Physical Review A, B, C, D, Physical Review Letter*, and *Reviews of Modern Physics*. APS journal paper citation trajectories have been extensively studied in prior literature (Redner, 2005; Wang, Song, & Barabási, 2013; Colavizza & Franceschet, 2016). We only include original research papers labeled as "article" and exclude other document types such as "review" or "note." There are a total of 4023 research articles, and their cumulative citations in the first 30 years after publication, i.e., between 1980 and 2009, are retrieved from the Web of Science (WoS). Since a sufficient amount of citations are required for reliable modeling of the citation trajectories (Aversa, 1985; Wang et al., 2013; Colavizza & Franceschet, 2016), we decide to focus on papers with at least 30 citations in the first 30 years after publication. The resulting dataset consists of 1699 papers. For a robustness check, we tested two different thresholds, specifically, we replicated the analysis using two different datasets: (1) papers with at least 20 total citations and (2) papers with at least 50 total citations. We obtained consistent clustering results (*Appendix B*).

Insert Fig. 2 here

There is considerable variation in individual papers' citation trajectories in our dataset. *Fig. 2* plots the citation trajectories of four selected papers. Three of them loosely resemble the three general types labeled by Costas et al. (2010) as *flash-in-the-pan* (red curve), *normal document* (blue curve), and *delayed document* (purple curve). The *normal document* (blue) follows the "typical" citation aging pattern, where the citations gradually increase and then decrease over time. The *flash-in-the-pan* (red) has relatively faster citation rising and declining processes, while the *delayed document* (purple) has relatively slower citation rising and declining process. All these three types follow the "typical" pattern of citation trajectory, although they vary in the general speed of citation ageing. However, the green curve has a continually rising annual citation curve without a declining stage during the first 30 years after being published. The green curve in *Fig. 2* illustrates the annual and cumulative citations of the paper in *Physical Review Letters* entitled "Ground State of the Electron Gas by a Stochastic Method" coauthored by Ceperley and Alderby, which is the most cited paper up to year 2009 among all papers published in 1980 in APS journals.

In addition, *Fig. 2* suggests a high level of uncertainty in paper citations and the difficulty of using short-term citations to predict long-term citations. Specifically, while *evergreen* papers would eventually become extremely highly cited, their citations in the first few years are not necessarily very large. Moreover, the distribution of the 30-year cumulative citation counts is highly skewed: 10 papers (0.6%) in our sample have citations greater than 1000, 50 papers (2.9%) greater than 400, and 1244 papers (73.2%) fewer than 100. This implies that the distribution of papers across different types of citation trajectories might also be uneven.

## 4. Methodology

The objective of this paper is to empirically uncover general types of citation trajectories based on the observed paper citation time series and examine whether *evergreens* constitute a general type of citation trajectory. The main idea is to use functional principal component analysis and Poisson regression to model citation trajectories. This approach allows us to conduct dimension reduction, that is, to characterize the vector of 30-year citation counts of a paper by a smaller number of parameters derived from our model. Subsequently these parameters can be used as inputs for the *K*-means cluster analysis for uncovering general types of citation trajectories.

### 4.1. Functional Poisson regression model

We develop a nonparametric model for the cumulative citations based on functional Poisson regression. The nonparametric approach does not impose any theoretical assumptions on the mechanisms underlying the citation process but lets the data speak for themselves. We adopt this explorative approach in order to better understand divergent citation patterns in real life.

For each paper $i = 1, …, N$, $x_i(j)$ denotes the observed cumulative number of citations for the *i*-th paper in year *j* after being published, where *j* is a discrete time variable (i.e., year) and $j = 1, …, T$. For notational convenience, we denote $x_i(0) = 0$. Our proposed functional Poisson regression model assumes that the observed cumulative citations $x_i(j)$ are the realization of a counting process $X_i(t)$ for the continuous time variable *t* and $0 \leq t \leq T$.

$$X_i(t) \sim Poisson(\mu_i(t)) \tag{1}$$

for $t \geq 0$, where the mean function $\mu_i(t)$ satisfies

$$\sqrt{\mu_i(t)} = \eta(t) + \sum_{v=1}^{\ell} \xi_{i,v} \phi_v(t) \tag{2}$$

and the function $\eta(t)$ and the basis functions $\phi_v(t)$ are smooth functions of *t* that are the same for all papers. Their estimations will be further discussed later.

In Equation (2) we adopt a square-root transformation for the mean function $\mu_i(t)$. Note that for Poisson regression, or more generally Generalized Linear Models, there are two popular transformation for the mean $\mu_i(t)$ of the count data: log-transformation $\log(\mu_i(t))$ and square-root transformation $\sqrt{\mu_i(t)}$ (Nelder & Baker, 1972; McCullagh & Nelder, 1989). For any given basis functions $\phi_v(t)$, both transformation strategies have been widely used in the statistics literature, and which transformation is better depends on the specific application and dataset. In the context of this study, the square root transformation is preferable. As will be explained in Section 4.2, in our functional principal component analysis for deriving basis functions, we will approximate the Poisson distribution of citation counts by a Gaussian distribution via the square-root transformation, which allows us to take advantage of the rich literature of FDA for Gaussian distributed data (Rice & Silverman, 1991; Ramsay & Silverman, 2005). Therefore, adopting the square-root transformation strategy here for the Poisson regression matches the square-root transformation in functional principal component analysis and consequently yields a better fit to the data.

In addition, in standard principal component analysis, the number $\ell$ of basis functions is assumed to be relatively small, while the retained basis functions should be able to explain most information of the original data. Under our model in Equations (1) and (2), the goal is essentially to find an estimate $\hat{\mu}_i(t)$ that is a smooth version of $X_i(t)$ with certain correlation structure. In addition, it is also useful to think our proposed model as a dimension reduction, representing the $T$-dimensional cumulative citations of a paper by a $\ell$-dimensional vector of coefficients $\xi_{i,v}$. Subsequently, the problem of identifying general citation patterns can be reduced to the cluster analysis of the $\ell$-dimensional vector of coefficients.

### 4.2. Model parameter estimation

When fitting the functional Poisson regression model in Equations (1) and (2) to the observed cumulative citations $x_i(j)$ of the $N$ papers, we need to estimate two kinds of unknown quantities: the common basis functions $\eta(t)$ and $\phi_v(t)$ which are the same for all papers, and the paper-specific coefficients $\xi_{i,v}$ which are tailored for each paper individually. Clearly they are closely related, and there are no unique estimation methods. Here we propose to estimate them by using the functional principal component analysis method and Poisson regression, respectively.

Regarding the estimation of the common basis functions $\eta(t)$ and $\phi_v(t)$ in Equation (2), intuitively one should use information across all the observed $N$ papers. From the functional decomposition viewpoint, these basis functions can be any set of orthogonal bases, although some bases are more efficient than others. In the functional data analysis literature, the estimation of these basis functions has been well-studied for Gaussian distributed data, e.g., Rice and Silverman (1991) and Ramsay and Silverman (2005). Here we propose to adapt these prior methods to Poisson count data by exploring the close relationship between Poisson and Gaussian distributions. For a Poisson random variable $X$ with a large mean $\mu > 0$, a well-known fact is $\sqrt{X} \sim N(\sqrt{\mu}, 0.5^2)$ (Thacker & Bromiley, 2001). Note that the variance of $\sqrt{X}$ is approximately constant, and thus the square-root transformation of Poisson data is also referred to as the variance-stabilizing transformation in the statistical literature (Anscombe, 1948). Brown, Carter, Low, and Zhang (2004) also used the square-root transformation to establish the global asymptotic equivalence between Poisson process and Gaussian process.

In this paper we consider the square-root transformation of the count variable, $\sqrt{X_i(t)}$, so that the bases $\eta(t)$ and $\phi_v(t)$ in Equation (2) can be estimated by applying the rich functional data analysis literature to the "approximate Gaussian" data $\sqrt{X_i(t)}$, e.g., Rice and Silverman (1991) and Ramsay and Silverman (2005). Specifically, the square-root transformation of the observed citation counts for each paper can be modeled as being independent realizations of a stochastic process $Y(t) = \sqrt{X(t)}$, with mean $E\big(Y(t)\big) = \eta(t)$ and covariance function $\gamma(s,t) = \mathrm{cov}\big(X(s), X(t)\big)$. We assume that there is an orthogonal expansion (in the L2 sense) of $\gamma(s,t)$ in terms of eigenfunctions

$$\gamma(s,t) \ = \ \sum_{\nu=1}^{\infty} \lambda_\nu \phi_\nu(s) \phi_\nu(t) \tag{3}$$

Then, according to the Karhunen–Loève expansion theorem, a random citation curve $Y_i(t) = \sqrt{X_i(t)}$ can be expressed as

$$\sqrt{X_i(t)} \ = \ \eta(t) + \sum_{\nu=1}^{\infty} \xi_{i,\nu} \phi_\nu(t) \tag{4}$$

where the coefficients $\xi_{i,\nu}$ are uncorrelated random variables with mean 0 and variance $\mathrm{Var}\big(\xi_{i,\nu}\big) = \lambda_\nu$ (Rice & Silverman, 1991; Hall et al., 2006).

Therefore, functions $\eta(t)$ and $\phi_\nu(t)$ in Equation (4) are close related to the mean function and correlation function of the stochastic process $Y(t) = \sqrt{X(t)}$, and we will use them as the basis functions in Equation (2). The basis functions $\eta(t)$ and $\phi_\nu(t)$ in Equation (4) for Gaussian data can be estimated by spline smoothing and functional principal component analysis methods in Rice and Silverman (1991) and Ramsay and Silverman (2005).

After estimating the common basis functions $\eta(t)$ and $\phi_\nu(t)$, the next step is to estimate the coefficients $\xi_{i,\nu}$ in the standard Poisson regression model from observed raw citations $x_i(j)$. This can be done by maximum likelihood estimation for Poisson regression, which is implemented in many statistical packages. In our analysis, the estimation of the coefficients $\xi_{i,\nu}$ is done on a Windows 8 Laptop with Intel i7-4510U CPU 2.0GHz by using the *glm*() function in the free statistical software *R* (version 3.1.1).

*4.3.    Cluster analysis*

Given that the $N$ papers and their corresponding cumulative citation curves $x_i(j)$ can be represented as $N$ points in the $\ell$-dimensional space of coefficients ($\xi_{i,1}, ..., \xi_{i,\ell}$), we propose to conduct cluster analysis by applying the $K$-means clustering algorithm to the reduced $\ell$-dimensional coefficient space.  In addition, in this $\ell$-dimensional coefficient space, the coefficients $\xi_{i,v}$ in Equation (2) correspond to different basis functions $\phi_v(t)$ and vary considerably in scale.  Therefore, we first standardize coefficients $\xi_{i,v}$ by

$$\tilde{\xi}_{i,v} = \frac{\xi_{i,v} - \mu_v}{s_v} \tag{5}$$

where $\mu_v$ and $s_v$ are respectively the mean and standard derivation of the fitted $N$ coefficient values ($\xi_{1,v}, ..., \xi_{N,v}$), for each principal component $v = 1, ..., \ell$.

Subsequently, we define the *distance* between papers in term of citation trajectories as the Euclid distance of the standardized coefficients ($\tilde{\xi}_{i,1}, ..., \tilde{\xi}_{i,\ell}$) in the $\ell$-dimensional space, based on which we use the $K$-means clustering algorithm to cluster papers into $K$ different groups.  Given the explorative nature of this study, we experiment and compare clustering results for $K = 2, 3, 4, 5,$ and 6 clusters.

*4.4.    Summary of methodology*

Our proposed functional Poisson regression model for clustering paper citation trajectories can be summarized as follows.

- Given the $T$-year cumulative citation trajectories of $N$ papers $x_i(j)$ for $i = 1, 2, \ldots, N$, and $j = 1, 2, \ldots, T$, first derive the square-root transformed data, $y_i(j) = \sqrt{x_i(j)}$.

- Estimate the mean functions $\eta(t)$ and eigenfunctions $\phi_v(t)$ of the transformed data $y_i(j)$, using functional principal component analysis.

- Determine $\ell$, the number of eigenfunctions $\phi_v(t)$ to retain.

- For each individual paper $i$, use the mean functions $\eta(t)$ and $\ell$ eigenfunctions $\phi_v(t)$ as basis functions and fit a Poisson regression model to its observed cumulative citation trajectory $(x_i(1), x_i(2), \ldots, x_i(T))$. This yields, for each individual paper, the estimated coefficients $(\xi_{i,1}, \xi_{i,2}, \ldots, \xi_{i,\ell})$. Accordingly, the $T$-dimension vector of cumulative citations for paper $i$, $(x_i(1), x_i(2), \ldots, x_i(T))$, can be represented by its $\ell$-dimensional vector of coefficients, $(\xi_{i,1}, \xi_{i,2}, \ldots, \xi_{i,\ell})$.

- Standardize each coefficient $\xi_{i,v}$ by $\tilde{\xi}_{i,v} = \frac{\xi_{i,v} - \mu_v}{s_v}$, where $\mu_v$ and $s_v$ are the mean and standard derivation of the $N$ fitted coefficient values $(\xi_{1,v}, \xi_{2,v}, \ldots, \xi_{N,v})$, for each principal component $v = 1, \ldots, \ell$.

- Apply the $K$-means clustering algorithm to the standardized coefficients $\tilde{\xi}_{i,v}$ to group $N$ papers into $K$ clusters.

## 5. Results

This section reports the numerical results of applying our proposed model and method to our sampled 1699 APS journal papers.

### 5.1. Estimating basis functions

The basis functions $\eta(t)$ and $\phi_v(t)$ play an important role in our proposed model and method, and they are estimated in $R$ (version 3.1.1) using the codes of Ramsay, Hooker, and Graves (2009).

<div align="center">Insert Fig. 3 here</div>

*Fig. 3* plots the estimated mean curve $\eta(t)$ and its first derivative $\eta'(t)$. Here $\eta(t)$ and $\eta'(t)$ are closely related to the average cumulative citations and average annual citations over time, respectively. The estimated first derivative $\eta'(t)$ is positive but decreases over time. This is consistent with the "typical" citation pattern that the annual citations generally are the largest in early years and subsequently decline slowly.

<div align="center">Insert Fig. 4 here</div>

*Fig. 4* plots the estimated smoothing versions of the first four eigenfunctions $\phi_v(t)$. They correspond to the four largest eigenvalues of 299.86, 16.39, 2.17 and 0.65, and these four eigenfunctions account for 93.8%, 5.1%, 0.7% and 0.2% of the total variability, respectively. The shape of these eigenfunctions indicates how a paper's cumulative citation trajectory might deviate from the mean curve $\eta(t)$. Specifically, the first smoothed eigenfunction $\hat{\phi}_1(t)$ is positive and monotonically increasing. Therefore, if a paper has a positive coefficient on $\hat{\phi}_1(t)$, then this paper will have more citations than the average paper (i.e., the mean curve) across all years, and more importantly its advantage over the average paper magnifies over time. This observation is consistent with the well-known *cumulative advantage* or *preferential attachment* phenomenon in citations. The second smoothed eigenfunction $\hat{\phi}_2(t)$ is positive in early years but negative in late years. If a paper has a positive coefficient on $\hat{\phi}_2(t)$, then this paper would have relatively more citations in early years than an average paper but fewer citations in later years,

<div align="center">14</div>

displaying a relatively fast citation ageing process. The third and fourth smoothed eigenfunctions, $\hat{\phi}_3(t)$ and $\hat{\phi}_4(t)$, capture more fine-grained fluctuation in citation trajectories over time. Furthermore, they both exhibit a periodic pattern, suggesting that the highly or less cited feature can be cyclic.

### 5.2. Determining the number of eigenfunctions

A critical step of our analysis is to decide how many eigenfunctions to retain, for which there is still no standard procedure in the FDA literature (Wang, Chiou, & Mueller, 2015). The rule of thumb is to choose a reasonably small number $\ell$ of eigenfunctions that not only explain high percentage (e.g., 95% or 99%) of total variation but also have a good fit to the observed data. Therefore, we take into account both the total explained variability and the goodness of fit.

In terms of explained variability, the first one, two and three eigenfunctions together account for 93.8%, 98.9% and 99.6% of the total variability, respectively. According to the rule of thumb, that is, 95% or 99% of total variation to retain, we can choose $\ell = 2$ or 3.

Insert Fig. 5 here

We then examine the goodness of fit. *Fig. 5* evaluates the goodness of fit for the first $\ell = 2,3,4,5$ basis functions using the mean squared error (MSE) criterion. More precisely, results in *Fig. 5* are based on 10-fold cross validation: We randomly partition the 1699 papers into 10 subgroups, where 9 subgroups have 170 papers and the 10th subgroup has 169 papers. Of the 10 subgroups, a single subgroup is retained as the validation set for evaluating the goodness of fitting, and the remaining 9 subgroups are used as training data to fit our proposed functional Poisson regression model using the $\ell = 2,3,4,5$ basis functions. We repeat the process 10 times, with each of the 10 subgroups used exactly once as the validation data to calculate the mean squared error (MSE). The average mean squared errors from these 10 repetitions are plotted in *Fig.5*. Based on this graph, we can adopt a strategy similar to the Cattell's scree test, that is, search for the elbow point. It seems that the goodness of fit improves considerably when increasing $\ell$ from 2 to 3, while a further increase in $\ell$ only improves the goodness of fit marginally.

Therefore, we choose $\ell = 3$, partly because increasing $\ell$ from 2 to 3 brings the largest improvement in fitting performance and partly because the first three eigenfunctions contain 99.6% variability, which is sufficiently high.

### 5.3. Fitting individual paper models

Based on the estimated basis functions, we fit our proposed functional Poisson regression model to each individual paper in the dataset, following the procedure described in section 4.2. For evaluating the fitness of our model, we compare our model with a recently developed parametric model for individual papers' citations documented in Wang et al. (2013) (hereafter the WSB model). Wang et al. (2013) model paper citations by a Poisson process, specifically, the expected cumulative number of citations of the $i$-th paper in year $t$ ($t \geq 0$) is

$$m \left( \exp \left\{ \lambda_i \Phi \left( \frac{\log(t) - \mu_i}{\sigma_i} \right) \right\} - 1 \right) \tag{6}$$

where $\Phi(t)$ is the cumulative density function of the standard normal $N(0,1)$ random variable, $\lambda_i$, $\mu_i$, and $\sigma_i$ are three paper-specific parameters that describe the citation trajectory of the $i$-th paper, and parameter $m$ is a global constant for the average citations of all papers and is set at 30 in Wang et al. (2013).

For fitting individual paper models, the natural choice is to use the estimated basis functions $\eta(t)$ and $\phi_v(t)$ in section 5.1 directly to derive the estimated coefficients $\xi_{i,v}$ in Equation (2) for each individual paper (as will be implemented in the next subsection for clustering analysis). However, using this approach for comparing model fitting performance is unfair to the WSB model, because our functional Poisson regression model would have used the same dataset twice: One at the population level for estimating basis functions and the other at the individual paper level for estimating paper-specific coefficients on the common basis functions. However, the WSB model uses the data only once.

Therefore, for a relatively fair comparison of model fitting, we use the same 10-fold cross-validation as discussed in Section 5.2. Specifically, we randomly partition the 1699 papers evenly into ten subgroups. For papers in each subgroup, we fit our functional Poisson regression model using the 3 basis functions estimated from papers in all the other nine subgroups. For papers in each subgroup, we also fit the WSB model separately. Then we calculate the mean squared error (MSE) of the fit by our model and WSB model.

Insert Fig. 6 here

To assess the goodness of fit, we compare the distribution of residuals. In addition, we plot log MSEs instead of MSEs at the original scale, considering that the distribution of MSEs is highly skew. *Fig. 6* left panel plots the kernel densities of log MSEs. Our functional Poison regression model clearly has smaller MSEs, and the Wilcoxon sum rank test further suggests that the MSEs of our proposed functional Poisson regression model are stochastically smaller than those of the WSB model. In addition, *Fig. 6* right panel reports a scatter plot of log MSEs, which suggests that our proposed model fits most papers (i.e., points below the diagonal line) better than the WSB model.

It is important to note that this comparison is still to some extent unfair to the WSB model. We adopted a 10-fold cross-validation strategy and used a separate training set for estimating our basis functions, although the training set does not overlap with the testing set, they are sampled from the same big dataset and therefore still share many things in common. In addition, the WSB model is developed for predicting long-term citations, while the goal of our model is to have a parsimonious characterization of citation trajectories with satisfactory goodness of fit. Therefore, the WSB model would avoid overfitting, while our model would intentionally over-fit the data to certain degree. For the same reason, we opted for the original WSB model documented in Wang et al. (2013) for this comparison, instead of the WSB-with-prior model documented in Shen, Wang, Song, and Barabási (2014). The WSB-with-prior model incorporates a conjugate prior and thereby reduces the number of estimated parameters, for avoiding overfitting. Compared with the original WSB model, the WSB-with-prior model has a lower fitting power but a higher prediction power. In summary, based on the comparison results, we do not claim that our model is superior to the WSB model, especially when considering their different purposes and structures, but only conclude that our model does fit the data well.

### 5.4. Clustering paper trajectories

Using the estimated basis functions $\eta(t)$ and the first three $\phi_\nu(t)$ from the whole sample of 1699 papers as reported in subsection 5.1, we estimate coefficients $\xi_{i,\nu}$ in Equation (2) for each of the 1699 papers. These estimated coefficients $\xi_{i,\nu}$ are then standardized and used as inputs for the *K-* means clustering analysis. Given the explorative nature of this clustering analysis, we experiment with different number of clusters, ranging from two to six.

Insert Fig. 7 here

We first report results for four clusters. To illustrate characteristics of the identified four clusters, i.e., four general types of citation trajectories, we find the centers of each cluster in the 3-dimensional standardized coefficients spaces and then convert them back into the original paper citations space to derive central curves in terms of cumulative and annual citations (*Fig. 7*). The number of observations in each cluster is as follows: red (972 papers, that is, 57.2% of the whole sample of 1699 papers), blue (454 papers, 26.7%), purple (228 papers, 13.4%), and green (45 papers, 2.6%).

Both the red and blue curves in *Fig. 7* are consistent with previous clustering studies (Aversa, 1985; Costas et al., 2010; Colavizza & Franceschet, 2016), in the sense that the speed of citation aging is slow for some papers while relatively fast for others. However, the year of citation peak seems to be the same for both the red and blue curves, while the only difference is about the scale of the peak. Therefore, both red and blue curves might belong to the category of *normal* documents as labeled by Costas et al. (2010). We name the red curve as *normal-low* and the blue curve as *normal-high.*

The purple curve, compared with both the red and blue ones, display a slower rising process, as well as a slower declining process after the citation peak. The timing of its citation peak is later than the red and blue ones. The scale of its citation peak is lower than the blue one but higher than the red one. In addition, its total number of 30-year citations is larger than both the red and blue ones.  The purple curve corresponds to the *delayed documents*, as labeled by Costas et al. (2010).

The most interesting curve in *Fig. 7* is the green one, which clearly demonstrates a continual rise in annual citations without declining within the 30-years period after publication.  We refer to this type of papers as *evergreens*, which were emphasized by Price (see Aversa (1985)) and Avramescu (1979) but were not identified by later cluster analyses (Aversa, 1985; Costas et al., 2010).  *Marathoners* in some specifications in Colavizza and Franceschet (2016) also display a continually increasing annual citation curve.  These *evergreens* appear to have fewer citations than the *normal-high* and *delayed documents* in the first few years after publications but clearly much more citations in the long run.  Furthermore, all other types (i.e., *normal-low*, *normal-high*, and *delayed documents*) still follow the "typical" citation trajectory, where a paper's annual citations rise to its peak shortly after publication and then slowly decline, although some types reach the citation peak higher or faster than others.  However, *evergreens* clearly violate this "typical" pattern, at least within the 30-year time window, which is much longer than the citation time window adopted in most bibliometric analyses.

Insert Fig. 8 here

Results for other choices of *K* are reported in *Fig. 8*.  On the one hand, decreasing *K* would miss some types of citation trajectories.  For example, the three-cluster result (*Fig. 8A3*) misses *delayed documents*, and the two-cluster result (*Fig. 8A2*) additionally misses *evergreens*.  On the other hand, increasing *K* from 4 to 5 or 6 does not uncover new types which are sufficiently distinct from the identified four types, and additional clusters in *Fig. 8A5-6* locate in a continuous space from fast to slow ageing, following the "typical" pattern.

In order to better evaluate the performance of our proposed clustering approach, we compare our proposed clustering method, which clusters citation trajectories based on the $\ell$-dimensional vector of standardized paper-specific coefficients $\tilde{\xi}_\nu^*$, with two alternative approaches, specifically, clustering based on (a) the $T$-dimensional vector of the raw annual citations (*raw annual method*) and (b) the $T$-dimensional vector of the proportion of annual citations (*proportion method*, i.e., normalized annual citations, the number of annual citations in each year divided by the number of total citations over the $T$ years). For the comparison of clustering results we focus on two aspects: the shape of the central curves and the distribution of papers across clusters.

Clustering results using the proportion method for $K = 2, \ldots, 6$ are reported in *Fig. 8B2-6*. Compared with our proposed method, the proportion method clusters papers more evenly across different clusters. In terms of the shape of the central curves, using $K = 4$ (*Fig. 8B4*) as an example, all four curves seem to reach their peak around the same time (while the green curve has an initial local peak at around the same time, followed by a decline and then starts rising again), although they display very different speed of citation declining. In addition, the speed of citation declining seems to be positively associated with the scale of the peak. For example, the blue curve has the highest peak and also the fastest citation decline after the peak. It is difficult to interpret the clusters. Maybe the red, purple, and blue curves can be labeled as *delayed document, normal document,* and *flash-in-the-pan* respectively, according to their speed of rising and declining, but the red one does not seem to have a later peak than the others. In addition, it is unclear how to interpret the green curve, it also exhibits a continual rise in annual citations (if we ignore the decline following the first local peak), similar to our identified *evergreens*. However, different from *evergreens*, the number of annual citations of the green type in *Fig. 8B4* is a small constant, and most papers in the green cluster have very limited number of total citations. One possible explanation is that this alternative approach uses the proportion of annual citations, which is very sensitive when a paper has a relatively small number of total citations.

Central curves of annual citations resulting from the clustering method based on raw annual citations are plotted in *Fig. 8C2-6*. The clustering result is dominated by the scale of citations, but does not reveal distinct features between different clusters in terms of the shape of citation curves. Take 4-cluster results (*Fig. 8C4*) as an example, 94.2% papers (red) have a moderate number of citations, 5.1% papers (purple) have even fewer citations, 0.6% papers (blue) have considerably more citations, and 0.1% papers (green) are extremely highly cited. Except the green curve, all others show a similar shape in the citation curve, and the difference between them is the scale of citations. Although this alternative approach also successfully identifies a small number of *evergreen* papers (i.e., the green curve), it misses out some true *evergreen* papers, that is, papers that are classified as *evergreens* by our proposed method but not by the raw annual method actually also exhibit a pattern of continual rise in annual citations. Thus, we conclude that clustering using raw annual citations is over-dominated by the scale of citations and is inadequate for capturing nuanced difference in the shape of citation trajectories.

## 5.5.    *Exploring characteristics of evergreens*

The clustering result clearly suggests the existence of *evergreens* as a general type of citation trajectory, in addition to previously documented *normal* and *delayed documents*. It also raises the question of how are four clusters differ from each other, in terms of various paper features, such as the number of authors and references. In addition, how do *evergreens* differ from others and can we predict whether a paper will become an *evergreen* paper based on its observed paper features?  To answer these questions, *Table 1* reports the means and medians of various paper features by clusters, and *Table 2* conducts nonparametric Wilcoxon rank sum tests for pairwise comparisons between four identified clusters.

Insert Table 1 here

Insert Table 2 here

The order of clusters from biggest to smallest in terms of the number of 3-year citations is: *normal-high*, *delayed*, *evergreen*, and *normal-low,* and pairwise differences are all significant except for between *evergreens* and *normal-low documents*.  On the other hand, the order from biggest to smallest in terms of 30-year citations is: *evergreens, delayed, normal-high,* and *normal-low*.  It is in line with *Fig. 7* that *evergreens* and *delayed documents* have fewer citations in the short run but much more citations in the long run, compared with *normal-high documents*, while *normal-low documents* are the majority of papers which have only relatively few citations throughout the whole time period.

In terms of the number of references, the only significant differences are that *delayed documents* have more references than *normal-low* and *normal-high* documents.  The order of clusters from biggest to smallest according to the number of pages is: *delayed*, *evergreens*, *normal-low*, *normal-high*, while *evergreens* are not significantly different from *delayed* or *normal-low*.  There seems to be a positive association between the number of pages and citation delay, in line with Wang, Thijs, and Glänzel (2015).

*Normal-high, normal-low, delayed,* and *evergreens*, following this order, have from the largest to the smallest number of authors, while the pairwise difference between *normal-low* and *delayed* *documents* is insignificant. E*vergreens* involve fewer institutes than *normal-high*, while *normal-high* have more institutes than *normal-low*. In addition, there are no significant differences between clusters in the number of countries. Interestingly, *evergreens* seem to have a small team size in terms of the number of authors or involved institutes. This observation reminds the thesis that breakthroughs are often delivered by "lone wolves" (Steinbeck, 1952), and future research is needed for a better understanding of this observation.

Insert Table 3 here

Insert Table 4 here

Focusing on *evergreens*, *Table 3* provides detailed characteristics of the *evergreen* cluster such as the average annual and cumulative citations in various years. Furthermore, we estimate two logistic regression models, testing whether *evergreens* can be identified based on the set of previously discussed paper features (*Table 4*). In both logistic regressions, the response for each paper is a binary variable: Whether the paper is an *evergreen* paper or not. The first logistic regression incorporates five *ex ante* explanatory variables which are determined upon publication, while the second adds two more *ex post* explanatory variables, specifically the log number of 3- and 30-year citations. Using the first model, if we classify papers as *evergreens* when the fitted probability is greater than 0.026, then the misclassification rate is 34.0%. If we classify papers with a fitted probability greater than 0.5 as *evergreens*, then the misclassification rate is 2.6%, but in this case no papers are classified as *evergreens*. The explanatory power of the first model is very low. Adding the citations improves the fitting performance. Using the second model, the misclassification rate is 8.5% and 1.9%, respectively, if we classify papers with a fitted probability greater than 0.026 and 0.5 as *evergreens*. The regression result confirms that *evergreens* tend to have relatively fewer citations in the short run but much more citations in the long run.  More importantly, the regression result suggests that we are not able to predict *evergreens* using readily-available *ex ante* paper features such as the number of authors or references, reflecting a high level of uncertainty in scientific impact.

## 6.  Discussion

This paper proposes a nonparametric functional Poisson regression model to describe citation trajectories of individual papers and combines our model with the *K*-means clustering algorithm for cluster analysis, using the coefficients of the eigenfunctions in our model.  Results suggest the existence of *evergreens* as a general type of citation trajectories.

This paper makes two methodological contributions.  First, we develop a functional data analysis method for discrete count data, by combining principal component analysis and Poisson regression, while the prior literature of functional data analysis is dominated by analyzing continuous data.  Second, this paper also demonstrates the usefulness of the functional data analysis for bibliometric studies.  Because it is a nonparametric approach and is designed for analyzing high-dimensional data, the functional data analysis can be a powerful tool for bibliometric analysis.

### 6.1.    Limitations and future research

This study has several limitations.  First, constrained by data availability, we cannot claim whether our observed *evergreen* papers will remain being (highly) cited in the future or will eventually become obsolete.  Although the latter is very plausible, the former is not entirely impossible.  Larivière, Archambault, and Gingras (2008) show that researchers have been relying on an increasingly old body of literature since the mid-1960s, so it is still possible that some classic pieces will never experience obsolesce or *obliteration by incorporation*, that is, becoming commonly known and integrated into the daily work in the field that it is no longer explicitly cited (Merton, 1983). Although we cannot draw a conclusive inference on the fate of our identified *evergreen* papers, the finding that a considerable number of papers assemble characteristics of *evergreens* in a 30-year time period is still very relevant for science and bibliometric studies, since most studies and evaluations use a shorter time window and assume a the "typical" citation trajectory. Second, this study uses a sample of journal articles in one field (i.e., physics) and one year (i.e., 1980), and accordingly has a limitation in terms of generalizability. Third, although our method can single *evergreens* out, it does not identify *sleeping beauties* in science (Van Raan, 2004; Ke, Ferrara, Radicchi, & Flammini, 2015). This is probably because *sleeping beauties* are very rare and therefore are difficult to identify in large scale statistical analyses (Colavizza & Franceschet, 2016).

There is plenty of room for improving our functional data analysis method for citation data, calling for further research. From the functional smoothing viewpoint, the cumulative citation curve must be non-decreasing. While our proposed fitting method yields non-decreasing fitted curves numerically for the cumulative citations of all 1699 papers in our dataset, it is important to develop a better estimation method that guarantees the non-decreasing property theoretically, e.g., using the monotone smoothing method developed in Ramsay (1998). From the cluster analysis viewpoint, we conduct unsupervised learning in our dataset and rely on prior literature and our domain knowledge on paper citation behavior, for assessing the classification results of different approaches. It will be useful to develop a more objective criterion for evaluating results of cluster analysis. In addition, we have some interesting observations, for example *evergreens* have a relatively small number of authors, more research is needed for better understanding what determines the citation trajectory of a paper. The regression model using readily-available paper feature for predicting *evergreens* has very poor performance, and it would interesting to investigate what kinds of intrinsic paper quality might predict whether a paper will become an *evergreen* in science.

## 6.2. *Implications*

Results of this paper have three important implications for bibliometric studies and research evaluations. First, our findings demonstrate that papers with similar citations in the short run may have completely different citation patterns in the long run. *Delayed documents* and *evergreens* receive fewer citations in the short run but more citations in the long run, compared with *normal documents*. This serves as a warning about the bias in the use of short-time-window citation counts in research evaluations.

Second, the observation of *evergreens* calls for more research on the "longevity" of citation impact, in addition to the aspect of "delay" emphasized in prior literature. Phenomena of *scientific prematurity* (Stent, 1972), *delayed recognition* (Garfield, 1980), and *sleeping beauties* (Van Raan, 2004) have been extensively studied in previous literature, which focus on the long time lag before a scientific contribution makes noticeable impact. On the other hand, *evergreens*, similar as the term of *marathoners* in Colavizza and Franceschet (2016), reminds the other important but understudied aspect of citation trajectory—unfading or long-lasting impact.

Third, *evergreens* also have implications for parametric models of citation trajectories. There is a strong interest in modelling citation trajectories, partly because it is a challenging scientific problem and partly because of the policy interest in predicting long-term citations. In a recent report published in *Science*, Wang et al. (2013) proposed a parametric nonhomogeneous Poisson process to model the citation trajectory of individual papers.  Although this model is elegant from the pure mathematical viewpoint, its predictive power is unsatisfactory, especially for those highly cited ones (Wang, Mei, & Hicks, 2014).  One possible explanation is that it assumes the "typical" citation trajectory, while *evergreens*, which are highly cited, do not follow this pattern. Results of our nonparametric analysis, in particular the observation of *evergreens*, cast doubt on this assumption and shed light on future parametric modeling of citation trajectories.

Reference

Anscombe, F. J. (1948). The Transformation of Poisson, Binomial and Negative-Binomial Data. *Biometrika, 35*(3/4), 246-254. doi:10.2307/2332343

Aversa, E. (1985). Citation patterns of highly cited papers and their relationship to literature aging: A study of the working literature. *Scientometrics, 7*(3-6), 383-389.

Avramescu, A. (1979). Actuality and obsolescence of scientific literature. *Journal of the American Society for Information Science, 30*(5), 296-303.

Baumgartner, S. E., & Leydesdorff, L. (2014). Group-based trajectory modeling (GBTM) of citations in scholarly literature: dynamic qualities of "transient" and "sticky knowledge claims". *Journal of the Association for Information Science and Technology, 65*(4), 797-811.

Besse, P., & Ramsay, J. O. (1986). Principal components analysis of sampled functions. *Psychometrika, 51*(2), 285-311.

Brown, L. D., Carter, A. V., Low, M. G., & Zhang, C.-H. (2004). Equivalence theory for density estimation, Poisson processes and Gaussian white noise with drift. 2074-2097. doi:10.1214/009053604000000012

Colavizza, G., & Franceschet, M. (2016). Clustering citation histories in the Physical Review. *Journal of Informetrics, 10*(4), 1037-1051. doi:http://dx.doi.org/10.1016/j.joi.2016.07.009

Costas, R., van Leeuwen, T. N., & van Raan, A. F. (2010). Is scientific literature subject to a 'Sell-By-Date'? A general methodology to analyze the 'durability'of scientific documents. *Journal of the American Society for Information Science and Technology, 61*(2), 329-339.

Garfield, E. (1980). Premature discovery or delayed recognition-Why. *Current Contents*(21), 5-10.

Glänzel, W., & Schoepflin, U. (1995). A bibliometric study on ageing and reception processes of scientific literature. *Journal of information Science, 21*(1), 37-53.

Hadjipantelis, P. Z., Aston, J. A., & Evans, J. P. (2012). Characterizing fundamental frequency in Mandarin: A functional principal component approach utilizing mixed effect models. *The Journal of the Acoustical Society of America, 131*(6), 4651-4664.

Hall, P., Müller, H.-G., & Wang, J.-L. (2006). Properties of principal component methods for functional and longitudinal data analysis. *The Annals of Statistics, 34*, 1493-1517.

Hoover, D. R., Rice, J. A., Wu, C. O., & Yang, L.-P. (1998). Nonparametric smoothing estimates of time-varying coefficient models with longitudinal data. *Biometrika, 85*(4), 809-822.

Ke, Q., Ferrara, E., Radicchi, F., & Flammini, A. (2015). Defining and identifying Sleeping Beauties in science. *Proceedings of the National Academy of Sciences, 112*(24), 7426-7431. doi:10.1073/pnas.1424329112

Larivière, V., Archambault, É., & Gingras, Y. (2008). Long-term variations in the aging of scientific literature: From exponential growth to steady-state science (1900–2004). *Journal of the American Society for Information Science and Technology, 59*(2), 288-296.

Leng, X., & Müller, H.-G. (2006). Classification using functional data analysis for temporal gene expression data. *Bioinformatics, 22*(1), 68-76.

Line, M. B. (1993). Changes in the use of literature with time-obsolescence revisited. *Library Trends, 41*(4), 665-684.

McCullagh, P., & Nelder, J. A. (1989). Generalized Linear Models, no. 37 in Monograph on Statistics and Applied Probability: Chapman & Hall.

Merton, R. K. (1983). Foreword. In E. Garfield (Ed.), *Citation indexing, its theory and application in science, technology, and humanities* (pp. xiii, 274 p.). Philadelphia, PA: ISI Press.

Nelder, J. A., & Baker, R. J. (1972). Generalized linear models. *Encyclopedia of statistical sciences*.

Ramsay, J. O. (1998). Estimating smooth monotone functions. *Journal of the Royal Statistical Society: Series B (Statistical Methodology), 60*(2), 365-375.

Ramsay, J. O., Hooker, G., & Graves, S. (2009). *Functional data analysis with R and MATLAB*: Springer Science & Business Media.

Ramsay, J. O., & Silverman, B. W. (2005). *Functional data analysis*: Springer.

Redner, S. (2005). Citation statistics from 110 years of Physical Review. *Physics Today, 58*(6), 49-54. doi:10.1063/1.1996475

Rice, J. A., & Silverman, B. W. (1991). Estimating the mean and covariance structure nonparametrically when the data are curves. *Journal of the Royal Statistical Society. Series B (Methodological)*, 233-243.

Rogers, J. D. (2010). Citation analysis of nanotechnology at the field level: implications of R&D evaluation. *Research Evaluation, 19*(4), 281-290.

Serban, N., Staicu, A. M., & Carroll, R. J. (2013). Multilevel Cross-Dependent Binary Longitudinal Data. *Biometrics, 69*(4), 903-913.

Shen, H.-W., Wang, D., Song, C., & Barabási, A.-L. (2014). Modeling and predicting popularity dynamics via reinforced Poisson processes. *arXiv preprint arXiv:1401.0778*.

Steinbeck, J. (1952). *East of Eden*. New York,: Viking Press.

Stent, G. (1972). Prematurity and uniqueness in scientific discovery. *Scientific American, 227*(6), 84-93.

Thacker, N. A., & Bromiley, P. A. (2001). The effects of a square root transform on a Poisson distributed quantity. *Tina memo, 10*, 2001.

van der Linde, A. (2009). A Bayesian latent variable approach to functional principal components analysis with binary and count data. *AStA Advances in Statistical Analysis, 93*(3), 307-333.

Van Raan, A. (2004). Sleeping beauties in science. *Scientometrics, 59*(3), 467-472.

Wang, D., Song, C., & Barabási, A.-L. (2013). Quantifying Long-Term Scientific Impact. *Science, 342*(6154), 127-132. Retrieved from http://www.sciencemag.org/content/342/6154/127.abstract

Wang, J.-L., Chiou, J.-M., & Mueller, H.-G. (2015). Review of functional data analysis. *arXiv preprint arXiv:1507.05135*.

Wang, J. (2013). Citation time window choice for research impact evaluation. *Scientometrics, 94*(3), 851-872. doi:10.1007/s11192-012-0775-9

Wang, J., Mei, Y., & Hicks, D. (2014). Comment on "Quantifying long-term scientific impact". *Science, 345*(6193), 149-149.

Wang, J., Thijs, B., & Glänzel, W. (2015). Interdisciplinarity and Impact: Distinct Effects of Variety, Balance, and Disparity. *Plos One, 10*(5), e0127298. doi:10.1371/journal.pone.0127298

Wu, S., Müller, H.-G., & Zhang, Z. (2013). Functional data analysis for point processes with rare events. *Statistica Sinica, 23*, 1-23.

Yao, F., Müller, H.-G., & Wang, J.-L. (2005). Functional data analysis for sparse longitudinal data. *Journal of the American Statistical Association, 100*(470), 577-590.

**Table 1**

Means and medians of selected variables by four identified clusters.

|  | normal-low | | normal-high | | delayed | | evergreen | |
|---|---|---|---|---|---|---|---|---|
|  | mean | median | mean | median | mean | median | mean | median |
| 3-year citations | 8.39 | 8.0 | 28.39 | 23.0 | 17.71 | 14.0 | 12.80 | 8.0 |
| 30-year citations | 56.59 | 47.0 | 106.70 | 79.0 | 216.96 | 150.5 | 602.67 | 355.0 |
| References | 26.28 | 22.0 | 27.58 | 20.0 | 32.68 | 26.5 | 30.22 | 22.0 |
| Pages | 8.50 | 7.0 | 8.19 | 4.0 | 13.93 | 10.0 | 12.64 | 9.0 |
| Authors | 3.11 | 2.0 | 4.65 | 3.0 | 2.96 | 3.0 | 2.42 | 2.0 |
| Institutes | 1.55 | 1.0 | 1.77 | 1.0 | 1.51 | 1.0 | 1.36 | 1.0 |
| Countries | 1.16 | 1.0 | 1.18 | 1.0 | 1.15 | 1.0 | 1.07 | 1.0 |

**Table 2**

Cluster pairwise comparison: Wilcoxon rank sum tests.

|  | evergreen – normal-low | evergreen – normal-high | evergreen – delayed | delayed – normal-low | delayed – normal-high | normal-high – normal-low |
|---|---|---|---|---|---|---|
| 3yr citations | 0.71 | **-7.77** | **-3.02** | **11.1** | **-11.81** | **29.7** |
| 30yr citations | **10.98** | **9.16** | **5.47** | **20.31** | **11.49** | **16.77** |
| References | -0.55 | -0.16 | -1.94 | **3.57** | **4.34** | -1.83 |
| Pages | 1.89 | **3.8** | 0.42 | **3.33** | **6.23** | **-6.54** |
| Authors | **-2.45** | **-3.66** | **-2.94** | 0.96 | **-2.23** | **4.14** |
| Institutes | -1.07 | **-1.84** | -1.04 | 0.05 | -1.55 | **2.24** |
| Countries | -0.82 | -0.77 | -0.83 | 0.13 | 0.15 | -0.04 |

Numbers are $Z$ statistics, bold numbers are significant at p < .05.

**Table 3**

*Evergreen* papers (*N* = 45)

| | Annual citations | Cumulative citations |
|---|---|---|
| | Mean ± Standard Err | Mean ± Standard Err |
| 3-year citations | 7.0 ± 1.2 | 12.8 ± 2.1 |
| 5-year citations | 8.9 ± 1.6 | 30.1 ± 5.0 |
| 10-year citations | 12.6 ± 2.8 | 83.9 ± 15.0 |
| 15-year citations | 18.9 ± 4.9 | 167.0 ± 34.3 |
| 20-year citations | 25.9 ± 7.2 | 283.8 ± 65.6 |
| 25-year citations | 31.5 ± 7.5 | 431.0 ± 101.0 |
| 30-year citations | 36.8 ± 10.3 | 602.7 ± 144.5 |

**Table 4**

Logit regression (*N* = 1699)

| | Being *evergreens* logit | |
|---|---|---|
| | (1) | (2) |
| 3-year citations (ln) | | -2.545 [0.358]*** |
| 30-year citations (ln) | | 3.700 [0.399]*** |
| References (ln) | -0.799 [0.283]*** | 0.357 [0.427] |
| Pages (ln) | 1.158 [0.278]*** | 0.475 [0.438] |
| Authors (ln) | -1.044 [0.453]** | -0.016 [0.610] |
| Institutes (ln) | -0.023 [0.801] | -0.161 [0.400] |
| Countries (ln) | -1.010 [1.230] | -2.665 [2.370] |
| Intercept | -1.616 [1.111] | -13.797 [2.438]*** |
| Residual deviance | 391.8 | 168.9 |
| Δ Residual deviance | | 222.9 *** |

Standard errors in brackets.  *** p<.01, ** p<.05, * p<.10.

# Appendix A. List of top cited APS journal papers

This appendix reports the list of the top 10 most cited APS journal papers in *Fig. 1*.

**Table 5**
Top 10 most cited APS journal papers

| | Title | Journal | Publication Year | Total citations |
|---|---|---|---|---|
| 1 | Development of the colle-salvetti correlation-energy formula into a functional of the electron-density | Physical Review B | 1988 | 57206 |
| 2 | Generalized gradient approximation made simple | Physical Review Letters | 1996 | 54662 |
| 3 | Density-functional exchange-energy approximation with correct asymptotic-behavior | Physical Review A | 1988 | 32095 |
| 4 | Efficient iterative schemes for ab initio total-energy calculations using a plane-wave basis set | Physical Review B | 1996 | 29064 |
| 5 | Inhomogeneous electron gas | Physical Review B | 1964 | 26005 |
| 6 | Special points for brillouin-zone integrations | Physical Review B | 1976 | 23806 |
| 7 | From ultrasoft pseudopotentials to the projector augmented-wave method | Physical Review B | 1999 | 21606 |
| 8 | Projector augmented-wave method | Physical Review B | 1994 | 20753 |
| 9 | Accurate and simple analytic representation of the electron-gas correlation-energy | Physical Review B | 1992 | 15122 |
| 10 | Abinitio molecular-dynamics for liquid-metals | Physical Review B | 1993 | 14908 |

Note: total citations are retrieved from Web of Science online interface on February 27, 2017.

# Appendix B. Robustness to alterative citation thresholds

Results reported in the main text are based on 1699 papers with at least 30 citations in the first 30 years after publication, we test whether the results are robust to alterative choices of citation thresholds. We replicated the analyses using two other datasets: (a) 2203 papers with at least 20 total citations and (b) 1078 papers with at least 50 total citations. Clustering results are consistent with what we report in the main text (*Fig. 9*).

Insert Fig. 9 here

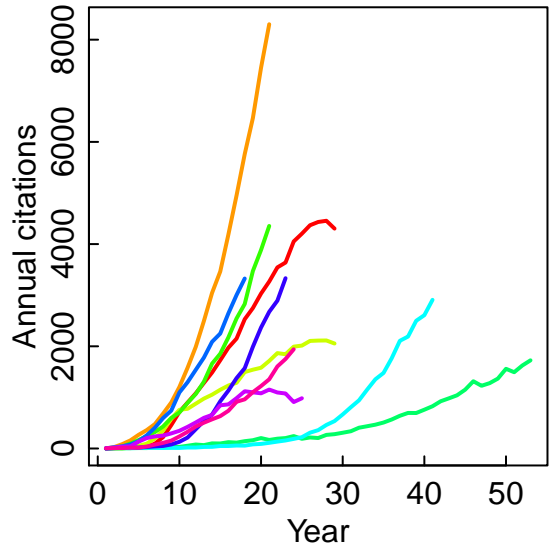

Figure 1: Annual citations of the top ten cited APS papers. One curve represents one paper, and details about these ten papers are reported in Appendix A.

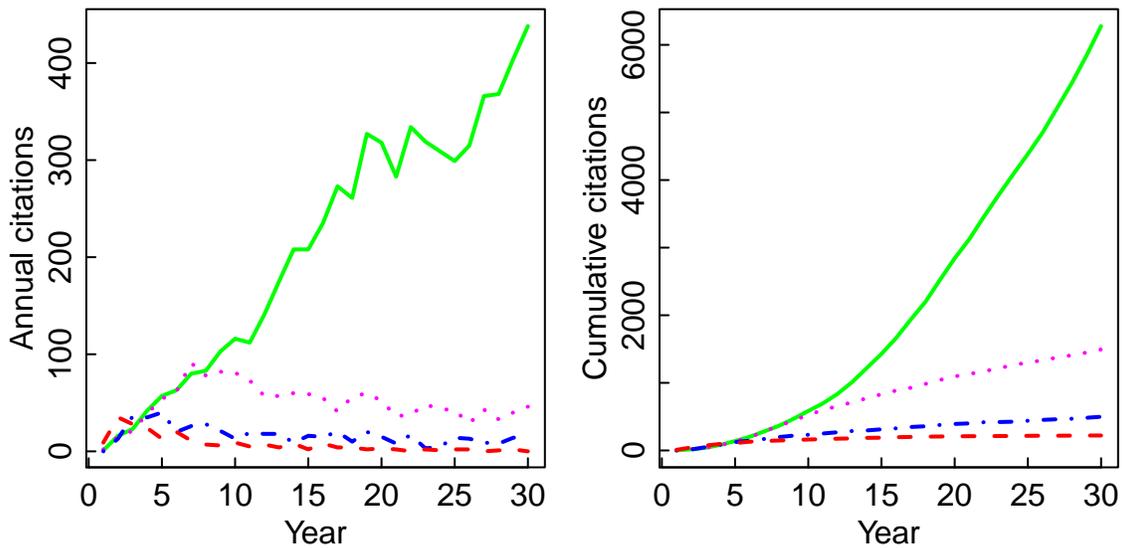

Figure 2: Annual and cumulative citations of four selected papers. One curve represents one selected paper. The red, blue, purple, and green curves correspond to *flash-in-the-pan*, *normal document*, *delayed document*, and *evergreen* respectively.

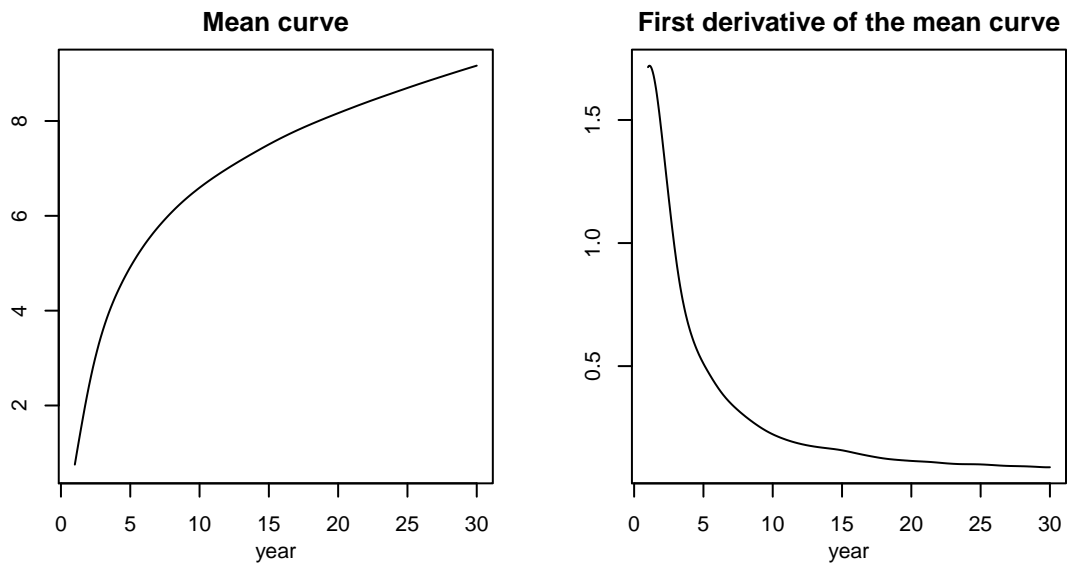

Figure 3: Mean function and its first derivative.

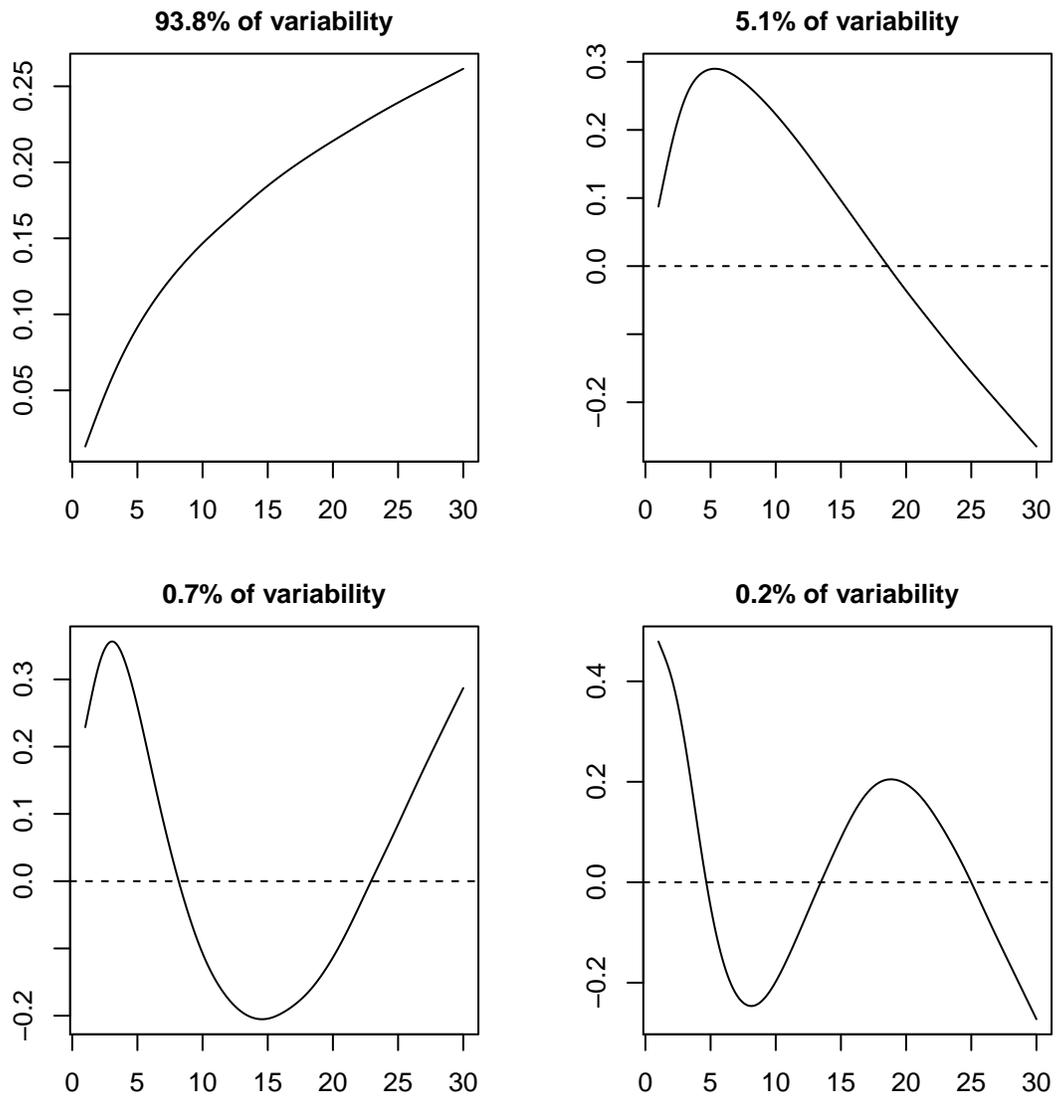

Figure 4: The first four eigenfunctions.

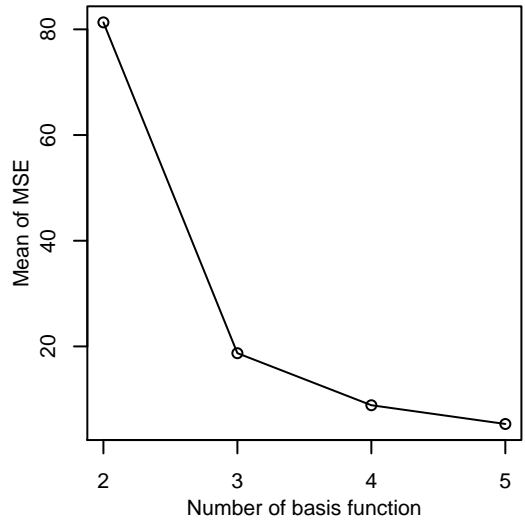

Figure 5: Determining the number of eigenfunctions.

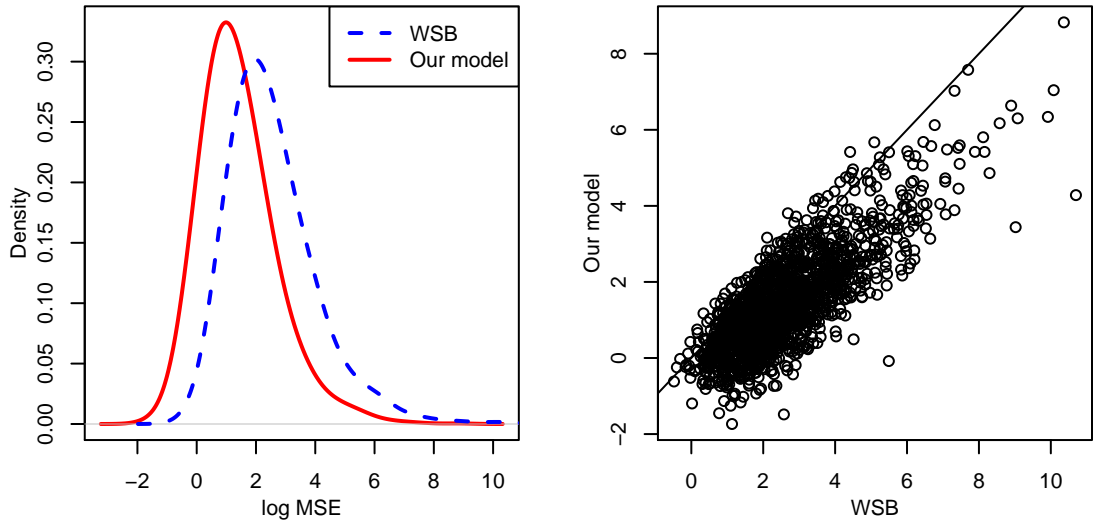

Figure 6: Goodness of fit. The left panel plots kernel densities of log MSEs. The right panel is a scatterplot, where one point represents one paper, and its $X$- and $Y$-axes are the log MSEs for the WSB model and our functional Poisson regression model respectively.

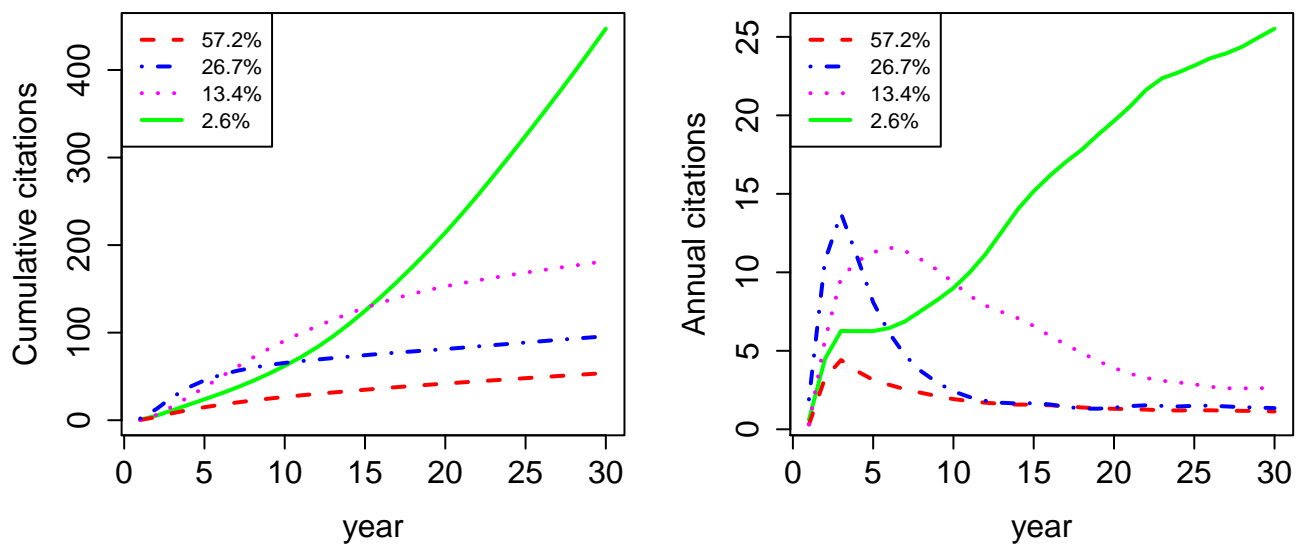

Figure 7: Clustering results: Four general types of citation trajectories. The red, blue, purple, and green curves represent *normal-low*, *normal-high*, *delayed*, and *evergreen* papers respectively.

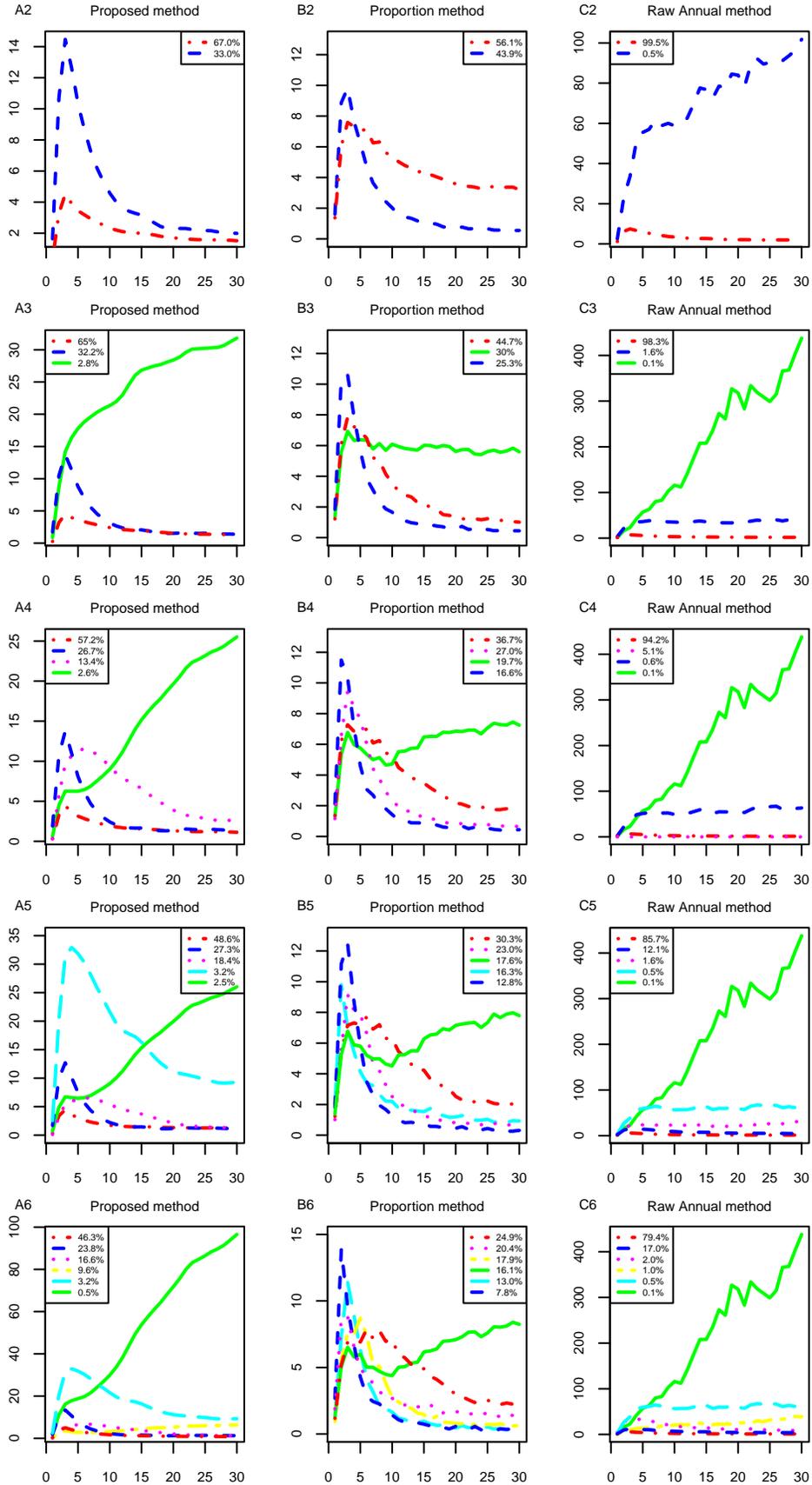

Figure 8: Clustering results, $K = 2$-6, three methods.

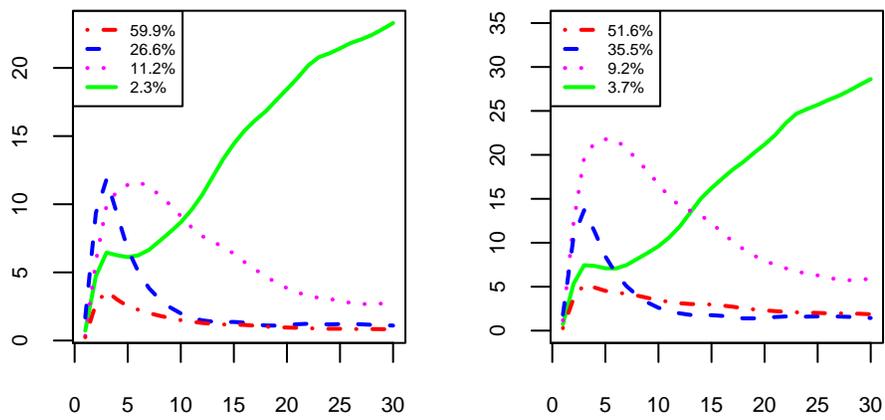

Figure 9: Clustering results: Alterative citation thresholds.